\documentstyle[11pt]{article}
\begin{document}
\begin{titlepage}
\null
\begin{flushright}
DFTT-68/92 \\
December, 1992
\end{flushright}
\vspace{1cm}
\begin{center}
{\Large\bf
Finite-size Effects in the Interface\\
of 3D Ising Model}
\end{center}
\vskip 1.3cm
\centerline{ M. Caselle, F. Gliozzi and S. Vinti}
\vskip .6cm
\centerline{\sl  Dipartimento di Fisica
Teorica dell'Universit\`a di Torino}
\centerline{\sl Istituto Nazionale di Fisica Nucleare,Sezione di Torino}
\centerline{\sl via P.Giuria 1, I-10125 Turin,
Italy\footnote{{\it email address}:~~~~~~Decnet=
(31890::CASELLE,GLIOZZI,VINTI)\\
\centerline{Internet=CASELLE(GLIOZZI)(VINTI)@TORINO.INFN.IT}}}
\vskip 2.5cm
\vskip 1.5cm

\begin{abstract}
The interface between domains of opposite magnetization in the 3D Ising
model near  the critical temperature displays universal finite-size
effects which can be described in terms of a gaussian model of
capillary waves. It turns out that these finite-size corrections
depend rather strongly on the shape of the lattice. This prediction,
which has no adjustable parameters,
is tested and accurately verified for various lattice shapes by means
of numerical simulations with a cluster algorithm. This supports also
a long-standing conjecture on the finite-size effects in Wilson loops of
Lattice Gauge Theories.

\end{abstract}

\normalsize
\end{titlepage}
\baselineskip=0.8cm
\setcounter{footnote}{0}

\newcommand{\eq}{\begin{equation}}
\newcommand{\en}{\end{equation}}
\newcommand{\eqa}{\begin{eqnarray}}
\newcommand{\ena}{\end{eqnarray}}
\newcommand{\spz}{\hspace{0.7cm}}
\newcommand{\lbl}{\label}
\newcommand{\lhi}{\hat\lambda_{i}}


\newcommand{\NP}[1]{Nucl.\ Phys.\ {\bf #1}}
\newcommand{\PL}[1]{Phys.\ Lett.\ {\bf #1}}
\newcommand{\NC}[1]{Nuovo Cim.\ {\bf #1}}
\newcommand{\CMP}[1]{Comm.\ Math.\ Phys.\ {\bf #1}}
\newcommand{\PR}[1]{Phys.\ Rev.\ {\bf #1}}
\newcommand{\PRL}[1]{Phys.\ Rev.\ Lett.\ {\bf #1}}
\newcommand{\MPL}[1]{Mod.\ Phys.\ Lett.\ {\bf #1}}
\newcommand{\IJMP}[1]{Int.\ J.\ Mod.\ Phys.\ {\bf #1}}
\newcommand{\JETP}[1]{Sov.\ Phys.\ JETP {\bf #1}}
\newcommand{\TMP}[1]{Teor.\ Mat.\ Fiz.\ {\bf #1}}


{\bf 1. Introduction}
\vskip .3cm
The properties of interfaces  in  3d statistical systems
have been a long standing subject of research.
In these last months there has been a renewed
interest on the subject, with a particular attention to the case
of the 3d Ising model,  partially
due to the improvement of Monte Carlo simulations: cluster algorithms
and multicanonical simulations allow now to compare theoretical
prediction on various physical quantities with reliable and precise
numerical estimates [1-6].
In particular, the interface above the roughening temperature behaves
like a free, fluctuating surface with a fixed topology
determined by the boundary conditions, so these new  algorithms
provide us with a powerful tool for testing our ideas on the behaviour
of fluctuating surfaces.

Another reason of interest on this subject is due
to the fact that in three dimensions ordinary spin models are related
through duality to lattice gauge theories (LGT). In particular the
physics of interfaces becomes in (LGT)'s the physics of Wilson loops,
and the string tension of the 3d Ising gauge model \cite{cfgpv,cfgv}
should have the same value of the interface tension of the 3d Ising spin
model, as it will be confirmed below.

In this letter we study the free-energy of the interface as a function
of its shape.
We will show that, using the capillary waves approach
and the semiclassical analysis of ref.~\cite{m}, it is possible to
infer a functional form of the interface free energy for
asymmetric lattices, which turns out to be
 in remarkable agreement with numerical
simulations. We will also discuss  implications of this result in the
context of LGT's.

This letter is organized as follows: in sect. 2 we discuss the capillary
wave approach in the case of  interfaces in  asymmetric lattices. In
sect. 3 we describe the simulation and the numerical results. In sect.4
we use duality to make contact with the 3d $Z_2$ gauge model and
 discuss the scaling behaviour of our results,  while the
last section is devoted to some concluding remarks.

\newpage
{\bf 2. Interface tension for asymmetric lattices}
\vskip .3cm
The interface between two equilibrium  phases above the
roughening temperature is essentially a critical system: it may be
freely translated through the medium and the long-wavelength,
transverse fluctuations in the position interface, i.e. capillary waves,
have  a small cost in energy (hence cannot be neglected in
calculations). They can be viewed as the Goldstone modes associated with
the spontaneous breaking of the transverse translational invariance
\cite{lu}.

The starting point of the capillary wave approach is the assumption that
these fluctuations are described by an effective Hamiltonian
proportional to the change they produce in  area of the interface
\eq
H/k_BT=\sigma\int_0^{L_1} dx_1
\int_0^{L_2}dx_2~ \left[\sqrt{1+\left(\frac{\partial h}{\partial x_1}\right)^2
+\left(\frac{\partial h}{\partial x_2}\right)^2}-1\right]~~,
\label{cw1}
\en
where the field $h(x_1,x_2)$ describes the interface displacement from
the equilibrium position as a function of the longitudinal coordinates
$x_1$ and $x_2$, $L_1~(L_2)$ is the
size of the lattice in the $x_1~(x_2)$ direction and $\sigma$ is the
(reduced) interface tension \footnote{Note that eq.(\ref{cw1}) coincides
with the Nambu-Goto string action in a special frame.}.

The interface free-energy increment due to interface
fluctuations (which we shall term from now on
capillary wave contribution to the interface tension) is then
given by

\eq
F_{cw}=-k_BT ~log~ tr~e^{-H/k_BT}~~,
\label{free}
\en
The Hamiltonian of eq.(\ref{cw1}) is  too difficult to be handled exactly.
However a crucial observation is  that this theory can be expanded in
the adimensional parameter $({\sigma L_1L_2})^{-1}$ and the leading order
term is the gaussian model, which turns out to be an accurate
approximation on the range of parameters used in current lattice
simulations. Then we replace eq.(\ref{cw1}) with the $\sigma L_1L_2
\to\infty$ limit $H \to H_G$

\eq
 H_G/k_BT=\frac{\sigma}{2}
\int_0^{L_1} dx_1 \int_0^{L_2} dx_2 ~ \left[\left(\frac{\partial h}
{\partial x_1}\right)^2
+\left(\frac{\partial h}{\partial x_2}\right)^2\right]~~.
\label{cw2}
\en
Within this approximation, the integration over $h$ implied in eq.
(\ref{free}) can be done exactly. The integral is divergent
but can be regularized using, for instance, a suitable generalization of
the Riemann $\zeta$-function regularization
(see $e.g.$~\cite{id}).
The answer  depends only on the geometrical properties of the boundary.
The case we are interested in is that of a rectangle of lengths $
L_1,L_2$ with opposite edges identified (hence a torus) due to the
periodic boundary conditions on the lattice. In this case the capillary
wave contribution turns out to be
\eq
F_{cw}(L_1,L_2)/k_BT=\log\left(\sqrt{-i\tau}|\eta(\tau)|^2\right)
+c\hskip0.5cm
;\hskip0.5cm \tau=i{L_1\over L_2}~~~,
\label{bos}
\en
\noindent
where $\eta$ denotes the Dedekind eta function:
\eq
\eta(\tau)=q^{1\over24}\prod_{n=1}^\infty(1-q^n)\hskip0.5cm
;\hskip0.5cmq=e^{2\pi i\tau}~~~,\label{eta}
\en
$c$ is an undetermined constant,
and we have assumed, without loss of generality, $L_1\geq L_2$.

Thus, the interface partition function $Z(L_1,L_2)$ may be written
in the form
\eq
Z(L_1,L_2)={\rm e}^{-\sigma L_1L_2}Z_{cw}~~ ,
\en
where the capillary wave contribution $Z_{cw}={\rm e}^{-F_{cw}/k_BT}$
can be expanded as follows

\eq
Z_{cw}(L_1,L_2)=\tilde C \sqrt{\frac{L_2}{L_1}} q^{-{\frac{1}{12}}}
      \left(1+q+2q^2+3q^3+5q^4+7q^5+11q^6+\dots\right)^2
\label{zcw}
\en
where $\tilde C$  is an undetermined constant and
$q=exp(-2\pi\frac{L_1}{L_2})$. This short expansion is largely
sufficient for an accurate numerical evaluation of this contribution in
the range of parameters we shall analyze.
Eq.(\ref{zcw}) is scale invariant, indeed it only
depends on the asymmetry parameter $ L_1/L_2$ and not on the
absolute value of $L_1$ or $L_2$. As a consequence, it cannot be
observed (since it only
gives a constant contribution) if one measures the interface tension on
$square$ lattices of different sizes, as it was done in
ref.~\cite{km,bhn,hp,h},
whose results are not affected by our calculation (and will be used in
our following analysis).

We want to stress that eq.(\ref{bos}) (or at least its dominant
contribution) is a rather well known result
both in conformal field theories and in  lattice
gauge theories, where it was first discussed by
L\"uscher, Symanzik and Weisz~\cite{lsw} as a
finite-size correction term
for Wilson loops (which, as we mentioned above, can be considered as
the dual problem of interfaces, see $e.g.$ \cite{id}).
To the best of our knowledge,
the first time it appeared in the context of interface physics is
in~\cite{gf}, where the leading term of eq.(\ref{bos}) was calculated
using the Epstein's function regularization. For a comparison with
numerical simulations we need the full form of eq.(\ref{bos}).

Let us conclude this section summarizing the various assumptions and
approximations involved in the  capillary wave approach
described above. First, we are
assuming that the surface separating the two phases has a well defined
localized shape. Second, we are neglecting bubbles and overhangs, namely
we are assuming that $h(x_1,x_2)$ is single-valued. Third, we are taking
small fluctuations. Fourth, we are assuming that the $h$ field does
not self-interact. The latter two approximations can be controlled (at
least in principle) by taking into account further orders in the
expansion parameter $({\sigma L_1L_2})^{-1}$ .

\vskip .5cm

{\bf 3. Monte Carlo simulation}
\vskip .3cm
We studied the 3d Ising model on an ordinary cubic lattice of size
$L_1\times L_2\times L_3$ and periodic boundary conditions in all
directions.
We performed our Monte Carlo simulation using the Swendsen-Wang cluster
algorithm~\cite{sw}.
$L_2$ and $L_3$ were kept fixed to the values $L_2=10$ and $L_3=120$
respectively, while $L_1$ ranged from $L_1=18$ to $L_1=32$.
As we shall see below these sizes are constrained by the requirement
$L_2<L_1\ll L_3$, and the chosen values of
$L_2$ and $L_3$ were the largest we could study with the CPU time we
had at our disposal. We shall call in the following $L_3$ the ``time ''
direction.

The Ising action is
\eq
S=-\sum_{\vec x,\mu}\sigma(\vec x) \sigma(\vec x+\hat\mu)
\en
 where the sum is over all sites $\vec x$ and all directions $\mu$ of the
lattice, and the spin variables can take the values $\sigma(\vec x)=\pm 1$.
The corresponding partition function is
\eq
Z=\sum_{\sigma(\vec x)=\pm1} e^{-\beta S}~~.
\en
The critical temperature of the model is estimated to be $\beta_c=0.221
654(6)$~\cite{betac}. We studied the model in the broken symmetry phase,
 at $\beta=0.2275$. This choice allowed us to make a direct comparison
with the data of~\cite{km}.
It is also well inside the scaling region of the dual gauge model,
as discussed in~\cite{cfgpv,cfgv}.

We  evaluated the interface tension  following the lines of
ref.\cite{km} (see also \cite{m,jjmmtw}) looking at the energy
splitting $E_0$ due to the
tunneling between the two vacua corresponding to the two  opposite
magnetizations of the system.
This energy splitting can be measured looking at the time-slice
correlations ( for this reason $L_3$ has been chosen much larger
than $L_1$ and $L_2$).

More precisely we define
\eq
S_t=\frac{1}{L_1L_2}\sum_{x_1=1}^{L_1}
\sum_{x_2=1}^{L_2} \sigma(x_1,x_2,t)~~,
\en
where $x_1,x_2,t$ are the coordinates in the $L_1,L_2,L_3$ directions
respectively.
We extracted from our simulations the averages $\langle S_0 S_t \rangle$.
Their  $t$-dependence is given by

\eq
\langle S_0 S_t \rangle~=~c_0 [e^{-tE_0}+e^{-(L_3-t)E_0}]~
+~c_1 [e^{-tE_1}+e^{-(L_3-t)E_1}] +\cdots
\label{fit}
\en
where the second exponent in each bracket is due to
 the periodic boundary conditions in the $L_3$ direction, and we have
explicitly written also the next to leading energy $E_1$, since it also
turns out to be important in the range of data we measure.

Finally  the interface tension can be extracted from $E_0$ by looking at
its volume dependence~\cite{fp}. In the case of a square section
$L_1=L_2\equiv L$ this is  given simply by:
\eq
E_0(L)=C~e^{-\sigma L^2}~~.
\label{fit1}
\en
This was the formula used in~\cite{km}, where a remarkable agreement
with numerical data was found.

Following the analysis of the previous section, the volume dependence of
$E_0$ on an asymmetric lattice (at the first order in the capillary wave
approximation) is given by

\eq
E_0(L_1,L_2)= Z_{cw}(L_1,L_2)~e^{-\sigma L_1L_2}
\label{fit2}
\en

where $Z_{cw}$ was defined in eq.(\ref{zcw}).
Since $Z_{cw}$ is scale invariant, and simply reduces to a
$L$-independent constant if $L_1=L_2\equiv L$, eq.(\ref{fit2})
is completely consistent
with eq.(\ref{fit1}), with the $same$ value of the interface tension
$\sigma$.

A simple way to check this picture is to evaluate the ratios
$R\equiv E_0(L_1,L_2)/E_0(\tilde L, \tilde L)$, with
$\tilde L\equiv \sqrt{L_1L_2}$. We get
\eq
R(L_1/L_2)\equiv \frac{E_0(L_1,L_2)}{E_0(\tilde L,\tilde L)}=
\left(\frac{\eta(i)}{\eta(i L_1/L_2)}\right)^2\sqrt{\frac{L_2}{L_1}}~~.
\label{ratio}
\en
The values of $E_0(\tilde L,\tilde L)$ are obtained extrapolating
eq.(\ref{fit1}), taking for $C$ and $\sigma$ the values quoted
in~\cite{km}, namely $\sigma=0.0150(1)$ and $C=0.188(4)$.

The ratios $R$, for various values of $L_1/L_2$, evaluated
theoretically ($R_{Th}$) and by Monte Carlo simulations ($R_{MC}$),
are presented in tab.1.
Error bars on $E_0$ were estimated with an ordinary jacknife procedure.
The same data are plotted in fig.1. The agreement between numerical
data and theoretical predictions is rather impressive: note that
the latter has no adjustable parameter, as it is shown by the
last member of eq.(\ref{ratio}).

Let us mention few non-trivial features of the simulation.
The approach outlined above allows very precise estimates of $E_0$ and
(with a lower precision) $E_1$, but, on the other side, time slice
correlations are affected by very strong cross-correlations. These must
be distinguished by the ordinary correlation in the Monte Carlo time,
which are almost completely eliminated by the use of the cluster
algorithm. Usually, cross-correlations are taken into account by weighting
the data in the fit with the inverted cross-correlation matrix, but in
this case cross-correlations were so strong that this method was almost
useless. We decided to eliminate cross-correlations directly in the
simulation by scattering the evaluation of time slice correlations in
Monte Carlo time (a method we had already used in the context of the Ising
gauge model~\cite{cfgpv,cfgv}, where similar, even if less severe,
problems occurred).

\noindent
Let us conclude this section with two remarks.

First, it is possible to show that the constant $c_0$ in
eq.($\ref{fit}$) is related to the spontaneous magnetization.
We checked that this relation is fulfilled in our samples.
Moreover, fitting our time slices correlations with eq.(\ref{fit}),
we always obtained the confidence levels between 50\% and 80\%.
Thus we are rather confident on the reliability of our estimates
of $E_0$.

Second, it is quite interesting to see how important is the
contribution of the  next to leading terms to the partition function
$Z_{cw}$. The difference between the
contribution of the whole $Z_{cw}$ given in eq.(\ref{zcw})
(full line in fig.1) and of its leading term $q^{-{\frac{1}{12}}}$
(dashed line in fig.1) is much bigger than our error bars.
Taking into account only this latter
 one would have a complete disagreement with the data and
would erroneously reject the capillary wave picture. The greatest part
of this difference is due to the square root term of eq.(\ref{zcw})
(which is  generated by the zero modes) . This  lesson  is particularly
important in the dual context of Lattice Gauge Theories, where
 it is a common habit to take into account (if any)
only the leading term of the L\"uscher-Symanzik-Weisz contribution.
It could well be that some of the scaling violations observed in recent
simulations of LGT could be due to this~\footnote{ Indeed
it has been shown in~\cite{cfgpv,cfgv} that  the
string tension extracted by Wilson loops (Polyakov lines)
taking into account the whole string corrections shows, as expected,
 a good scaling behaviour.}.

\vskip .5cm

{\bf 4. Scaling and duality}
\vskip .3cm
The 3d Ising model is related through duality to the 3d $Z_2$ gauge
model.
Let us call $\tilde\beta\equiv \frac{1}{k_B\tilde T}$ the inverse
temperature of the gauge model. $\beta$ and $\tilde \beta$ are related
by
\eq
\beta=-\frac{1}{2}log~tanh~\tilde\beta
\hskip 1cm
\tilde \beta=-\frac{1}{2} log~tanh~\beta~~~.
\label{dual}
\en
The critical point $\beta_c=0.2216\cdots$ is mapped by duality in
$\tilde\beta_c=0.7614\cdots$.

As we mentioned above, the capillary wave corrections have their
parallel in the dual context of lattice gauge theory in the so
called string corrections, namely the sum over all the possible
fluctuations of the flux tube (thought as a string-like object)
joining a quark-antiquark pair. Roughly speaking, the time world-lines
of a quark-antiquark pair form a Wilson loop (or a pair Polyakov
lines in the case of finite temperature LGT's); then, in order to
evaluate the string corrections one has to sum over all the possible
fluctuations of the surface bordered by the Wilson loop (or by the
Polyakov lines). It is thus apparent that the only difference between
the present analysis and those performed in LGT's is in the boundary
conditions. In the case of Wilson loops one has fixed  boundary
conditions, hence a rectangular geometry. In the case of Polyakov line
correlations one has fixed boundary conditions in one direction and
periodic in the other, hence a cylindric geometry. Finally in the
present analysis we have a toroidal geometry. In all these cases the
asymptotic string (or capillary wave) contributions can be evaluated
exactly (we refer to~\cite{cfgpv} and to~\cite{cfgv} for the discussion
of the rectangular and cylindric cases respectively).

In~\cite{cfgpv} we evaluated the string tension of the Ising gauge model
for various values of $\tilde\beta$ looking at the expectation value of
Wilson loops. These data are collected in tab.2 and plotted in fig.2
together with those on the interface tension taken from
ref.s~\cite{km,bhn,h}. One can see that the two sets of data are perfectly
compatible. The only difference is that those coming
from the gauge model have bigger errors  because
the algorithms recently introduced for the calculation of the interface
tension allow, with the same CPU time, an impressive improvement in
precision. Notice that in the data taken from~\cite{cfgpv}, string
corrections were already taken into account, hence the remarkable
numerical coincidence of string and interface tension yields further
support to the reliability of the capillary wave (or string in LGT's)
approximation. Fitting the data of ref.~\cite{cfgpv} according to
the law $\sigma(\beta)=\sigma_0(1-\beta_c/\beta)^{2\nu}$, with the fixed
value of $\nu=0.63$~\cite{nu}, we obtain $\sigma_0=1.52(2)$ ($\chi^2/
d.o.f.\sim 0.8$) in perfect agreement with the results of [1-6].

It has to be noticed that in our analysis of Wilson loops \cite{cfgpv,cfgv,lat}
we used a modified form of the gaussian model, where the field $h$
is compactified on a circle of suitable radius. As a consequence, the
partition function has a form very different from that given in
eq.(\ref{zcw}). Surprisingly enough, in the range $1\leq L_1/L_2\leq5$,
the values of the ratios defined in eq.(\ref{ratio}) are almost the same
of those calculated with eq.(\ref{zcw}).
\vskip .5cm

{\bf 5. Conclusions}
\vskip .3cm

In this letter we have shown that the finite-size effects in  the
interface  of 3d Ising
model is well described by the capillary wave theory. We have
shown that a simple, precise way to check this description is by looking
at the interface tension on asymmetric lattices. The agreement between
numerical data and theoretical prediction is quite remarkable once the
whole capillary wave contribution has been taken into account. In
particular we have shown that a relevant role is played by the
subdominant logarithmic correction contained in $F_{cw}$.

It would be interesting to obtain our results with other simulation
techniques like the multi-magnetic approach~\cite{bhn} or the cluster
update of boundary conditions proposed in~\cite{h}.
In this way one would probably obtain higher precision and it would
be possible to see higher order corrections.
These could be calculated perturbatively by expanding the theory
in the parameter $({\sigma L_1 L_2})^{-1}$ . The first correction is
the quartic term coming from the expansion of the square root in
eq.(\ref{cw1}) (see $e.g.$~\cite{df}).

It would be also interesting to repeat the same analysis with other
models to see how universal are the predictions of the capillary wave
approximation. Indeed it is well known that the roughening phenomenon,
which can be described within the same approach, shows an high degree of
universality.


\vskip .5 cm
\hrule
\vskip .4 cm
\centerline{\sl Figure Captions}
\begin{description}

\item{Fig.1)}
The values of the ratios $R$ defined by eq.(\ref{ratio}) are
plotted: the continuous line and the dotted line being respectively
calculated using  eq.(\ref{zcw}) and taking into account
only the $q^{-{\frac{1}{12}}}$ term of the same equation.
The ratios evaluated by Monte Carlo simulations are also
plotted with error bars.
Note that the theoretical prediction has no free parameters.

\item{Fig.2)}
The interface (string) tension is plotted for various
values of $\beta$. The data with error bars refer to the string tension
of Wilson loops in the Ising gauge model and are taken
from~\cite{cfgpv}. The other data refer to the interface tension in the
Ising spin model and are taken from ref.s~\cite{km,bhn,h},
their error bars are not reported since are smaller than the plotted
symbols. The continuous line is the fit of the data of ref~\cite{cfgpv}
 only, with the law
$\sigma(\beta)=\sigma_0(1-\beta_c/\beta)^{1.26}$.

\end{description}

\vskip .2 cm
\hrule
\vskip .4 cm

\centerline{\sl Table Captions}
\begin{description}

\item{Tab.I}
The same data of fig.1. are reported:
in the first column the ratio $L_1/L_2$, in
the third and fourth columns the corresponding ratios from
the Monte Carlo ($R_{MC}$) and theoretical ($R_{Th}$) evaluation,
while the values obtained taking into account only
the $q^{-{\frac{1}{12}}}$ term is given in the last
column.
In the second column are written the values of the energy splitting
$E_0$ that we measured on the corresponding asymmetric lattices.

\item{Tab.II}
The same data of fig.2. are reported:
in the first column the values of $\beta$, in the second the
corresponding values of the interface (string) tension and in the third
the values of $\sigma_0$ obtained using the relation
$\sigma_0=\sigma(\beta)(1-\beta_c/\beta)^{-1.26}$.
In the last column are reported the ref.s from which the data are taken
and the simulation techniques adopted: WL=Wilson loops, MM=Multimagnetic
simulation, CUBC=Cluster update of boundary conditions, ES=energy
splitting.

\end{description}

\vskip .2 cm
\hrule
\vskip .8 cm
\centerline {\bf Tab. I}
$$\vbox {\offinterlineskip
\halign  { \strut#& \vrule# \tabskip=.5cm plus1cm
& \hfil#\hfil & \vrule# & \hfil# \hfil
& \vrule# & \hfil# \hfil
& \vrule# & \hfil# \hfil &
& \vrule# & \hfil# \hfil &\vrule# \tabskip=0pt \cr \noalign {\hrule}

&& $x\equiv L_1/L_2$ && $E_0(L_1,L_2)$  && $R_{MC}(x)$  && $R_{Th}(x)$
&& $e^{{\pi\over6}(x-1)}$ & \cr \noalign {\hrule}
&& $18/10$ && $0.01439~(26)$ && $1.14~~(6)$ &&
$1.13$ && $1.52$ & \cr \noalign {\hrule}
&& $20/10$ && $0.01099~(23)$ && $1.17~~(7)$ &&
$1.19$ && $1.69$ & \cr \noalign {\hrule}
&& $23/10$ && $0.00797~(23)$ && $1.33~(10)$ &&
$1.30$ && $1.97$ & \cr \noalign {\hrule}
&& $26/10$ && $0.00584~(23)$ && $1.53~(13)$ &&
$1.43$ && $2.31$ & \cr \noalign {\hrule}
&& $28/10$ && $0.00396~(44)$ && $1.40~(22)$ &&
$1.53$ && $2.57$ & \cr \noalign {\hrule}
&& $32/10$ && $0.00289~(32)$ && $1.87~(31)$ &&
$1.76$ && $3.16$ & \cr \noalign {\hrule}
}}$$
%
%
\centerline {\bf Tab. II}
$$\vbox {\offinterlineskip
\halign  { \strut#& \vrule# \tabskip=.5cm plus1cm
& \hfil#\hfil & \vrule# & \hfil# \hfil
  & \vrule# & \hfil# \hfil &
& \vrule# & \hfil# \hfil &\vrule# \tabskip=0pt \cr \noalign {\hrule}
&& $\beta$ && $\sigma$  && $\sigma_0$ && $meth.~\&~ref.$
& \cr \noalign {\hrule}
&& $0.2230$ && $0.00252~~(3)$ && $1.586~(19)$ && $CUBC$~~[6]
 & \cr \noalign {\hrule}
&& $0.2240$ && $0.00492~~(4)$ && $1.529~(12)$ && $CUBC$~~[6]
 & \cr \noalign {\hrule}
&& $0.2255$ && $0.00904~~(6)$ && $1.529~(10)$ && $CUBC$~~[6]
 & \cr \noalign {\hrule}
&& $0.2258$ && $0.0093~~(5)~$ && $1.442~(78)$ && $WL$~~[7]
 & \cr \noalign {\hrule}
&& $0.2269$ && $0.0135~~(5)~$ && $1.538~(57)$ && $WL$~~[7]
 & \cr \noalign {\hrule}
&& $0.2270$ && $0.01293~(17)$ && $1.457~(19)$ && $MM$~~[3]
& \cr \noalign {\hrule}
&& $0.2275$ && $0.0150~~(1)~$ && $1.508~(10)$ && $ES$~~[2]
 & \cr \noalign {\hrule}
&& $0.2281$ && $0.0177~~(5)~$ && $1.575~(45)$ && $WL$~~[7]
 & \cr \noalign {\hrule}
&& $0.2293$ && $0.0212~~(5)~$ && $1.537~(37)$ && $WL$~~[7]
 & \cr \noalign {\hrule}
&& $0.2305$ && $0.0244~~(5)~$ && $1.484~(31)$ && $WL$~~[7]
 & \cr \noalign {\hrule}
&& $0.2317$ && $0.0292~~(6)~$ && $1.523~(37)$ && $WL$~~[7]
 & \cr \noalign {\hrule}
&& $0.2320$ && $0.03140~(14)$ && $1.575~~(7)$ && $MM$~~[3]
& \cr \noalign {\hrule}
&& $0.2327$ && $0.0328~~(2)$ && $1.525~(10)$ && $ES$~~[2]
 & \cr \noalign {\hrule}
}}$$

\begin{thebibliography}{99}

\bibitem{gprs} H.Gausterer, J.Potvin, C.Rebbi and S.Sanielevici,
preprint BU-HEP-92-16.

\bibitem{km} S.Klessinger and G.\"Munster, preprint MS-TPI-92-13, May
1992;

\bibitem{bhn} B.Berg, U.Hansmann and T.Neuhaus, preprint BI-TP 92/10,
June 1992;

\bibitem{mt} H.Meyer-Ortmanns and T.Trappenberg, J. Stat. Phys. {\bf 58}
(1990) 185;

\bibitem{hp} M.Hasenbusch and K.Pinn, preprint MS-TPI-92-24,
September 1992;

\bibitem{h} M.Hasenbusch, preprint KL-TH 16/92, September 1992;


\bibitem{cfgpv} M.Caselle, R.Fiore, F.Gliozzi, P.Provero and S.Vinti,
\IJMP{6A} (1991) 4885

\bibitem{cfgv} M.Caselle, R.Fiore, F.Gliozzi and S.Vinti,
 preprint DFTT 23/92, to be published on Int. J. Mod. Phys.

\bibitem{m} G.M\"unster, \NP{B340} (1990) 559

\bibitem{lu}
M.L\"uscher, \NP{B180} (1981) 317

\bibitem{id} J.M.Drouffe and C.Itzykson, {\sl Statistical Field Theory},
Cambridge University Press (1989)

\bibitem{lsw}
M.L\"uscher, K.Symanzik and P.Weisz, \NP{B173} (1980) 365

\bibitem{gf} M.Gelfand and M.Fisher, Physica {\bf A166} (1990) 1


\bibitem{jjmmtw} K.Jansen, J.Jersak, I.Montvay, G.M\"unster,
T.Trappenberg and U.Wolff, \PL{B213} (1988) 203

\bibitem{df} K.Dietz and T.Filk, \PR{D27}, (1983) 2944

\bibitem{betac} G.S.Pawley et al., \PR{B29} (1984) 4030

M.N. Barber et al.,\PR{B32} (1985) 1720

\bibitem{sw} R.H.Swendsen and J.S. Wang, \PRL{58} (1987) 86

\bibitem{fp} M.E.Fisher and V.Privman, Journal of Stat. Phys. {\bf 33}
(1983) 385.

\bibitem{nu} N.A.Alves, B.A.Berg and R.Villanova, \PR{B41} (1990) 383

\bibitem{lat} M.Caselle, F.Gliozzi and S.Vinti, Talk given at
Lattice'92, Amsterdam, to be published on Nucl Phys.{\bf B}, Proc. Supp.


\end{thebibliography}
\end{document}